\documentclass[aps,prl,twocolumn]{revtex4-1}  

\usepackage{bbm,graphicx,bm}

\def\ii{{\rm i}}

\def\ket#1{| {#1} \rangle}
\def\bra#1{\langle {#1} |}
\def\tr#1{{\rm tr}{{#1}}}

\def\tit#1{{\em #1},}
\def\etal#1{#1}

\begin{document}

\title{Coexistence of diffusive and ballistic transport in a simple spin ladder}

\author{Marko \v Znidari\v c}
\affiliation{
Physics Department, Faculty of Mathematics and Physics, University of Ljubljana, Ljubljana, Slovenia}

\date{\today}

\begin{abstract}
We show that in a nonintegrable spin ladder system with the XX type of coupling along the legs and the XXZ type along the rungs there are invariant subspaces that support ballistic magnetization transport. In the complementary subspace the transport is found to be diffusive. This shows that (i) quantum chaotic systems can possess ballistic subspaces, and (ii) diffusive and ballistic transport modes can coexist in a rather simple nonintegrable model. In the limit of an infinite anisotropy in rungs the system studied is equivalent to the one-dimensional Hubbard model.
\end{abstract}



\maketitle

Transport properties of simple spin ladders are interesting for two reasons. On one hand they are realized in a number of materials~\cite{review,dagotto:99,hess}, on the other they serve as model systems on which theoretical ideas can be tested. For instance, the Hubbard model -- a paradigmatic model of strongly correlated electrons -- is equivalent to a ladder system. One of the most actively investigated areas of statistical physics in recent years is nonequilibrium properties of strongly correlated systems. In particular, there is a long quest to understand transport properties of systems from first principles, e.g.~\cite{lebowitz}. For recent studies of transport in spin ladders see~\cite{sachdev,alvarez,meisner,orignac,zotos,boulat}. With rapidly progressing experimental cold atoms techniques~\cite{Bloch12}, transport properties of Fermi gases~\cite{Zweirlein11} as well as of the Hubbard model~\cite{Schneider12} have actually been measured. Perhaps the simplest question, is a given model diffusive or ballistic, seems to be for most models too difficult to rigorously answer, even if the system is integrable, an example being for instance the gapped one-dimensional Heisenberg model, see, e.g.~\cite{Affleck}. A powerful method to prove ballistic transport is by bounding the time-averaged current autocorrelation function using constants of motion, the so-called Mazur's inequality~\cite{mazur,Zotos1997}. Because quantum integrability is usually defined~\cite{Qintegrability} by the existence of an infinite set of local conserved quantities it often leads to ballistic transport. In fact, all proved ballistic systems are integrable and possess either local~\cite{Zotos1997} or quasilocal~\cite{prosen} conserved quantities that have nonzero overlap with the current. Based on this one is tempted to conclude that ballistic transport is possible only in integrable systems and that chaotic ones display diffusion. However, as we shall show, this widely held belief is not correct. Studying a class of spin ladder systems, which includes nonintegrable as well as integrable instances (the Hubbard model), we explicitly show the existence of ballistic transport. This provides a new mechanism of ballistic transport, different from the so-far known ballistic transport in integrable systems.

{\em XX-ladder.--} We shall study the so-called XX spin ladder composed of two coupled spin-$1/2$ chains in which the nearest-neighbor coupling along two chains (legs) is of the XX type, while the interchain coupling (rungs) is of the XXZ type. We shall show that the XX ladder, regardless of the value of two parameters $J$ and $\Delta$, possesses a number of ballistic invariant subspaces. That is, there exist subspaces of the total Hilbert space that are invariant under unitary evolution generated by $H$, and in which inhomogeneities spread out with a constant velocity, implying ballistic transport. In addition, we shall show that in the rest of the Hilbert space the transport is diffusive. The Hamiltonian is,
\begin{eqnarray}
\label{eq:XXladder}
&H&=\sum_{i=1}^{L-1} h^{||}_{i,i+1}+\sum_{i=1}^{L} h^{\perp}_{i},\\
&h^{||}_{i,i+1}&=\sigma_i^{\rm x} \sigma_{i+1}^{\rm x}+\sigma_i^{\rm y} \sigma_{i+1}^{\rm y}+\tau_i^{\rm x} \tau_{i+1}^{\rm x}+\tau_i^{\rm y} \tau_{i+1}^{\rm y},\nonumber \\
&h^{\perp}_{i}&=J(\sigma_i^{\rm x} \tau_{i}^{\rm x}+\sigma_i^{\rm y} \tau_{i}^{\rm y}+\Delta\, \sigma_i^{\rm z} \tau_{i}^{\rm z}).\nonumber
\end{eqnarray}
The model studied, Eq.~(\ref{eq:XXladder}), contains the integrable Hubbard model as a special limit when $J \to 0$ and $J\Delta \to \mathrm{const.}$~\cite{shastry86}. Let us first briefly mention standard symmetries of the XX ladder. Total magnetization along the ${\rm z}$-axis, $Z=\sum_{j=1}^L \sigma_j^{\rm z}+\tau_j^{\rm z}$, is conserved. There are also two lattice symmetries, namely, a parity $P_{\rm x}$ that exchanges sites $j$ and $L+1-j$, and a parity $P_{\rm y}$ that exchanges the two chains. In the sector with $Z=0$ there is an additional spin-flip symmetry corresponding to the transformation $F=\prod_j \sigma_j^{\rm x} \tau_j^{\rm x}$. There exist though additional symmetries that are a consequence of an XX-type interaction along the legs. Discussion is simple in the rung eigenbasis. On one rung the eigenvectors and eigenvalues are the singlet $\ket{\rm S}\equiv (\ket{01}-\ket{10})/\sqrt{2}$ with $E_{\rm S}=-(2+\Delta)J$, and three triplet states, $\ket{\rm T}\equiv (\ket{01}+\ket{10})/\sqrt{2}$ with $E_{\rm T}=(2-\Delta)J$, $\ket{\rm O}\equiv \ket{00}$ with $E_{\rm O}=J\Delta$ and $\ket{\rm I}\equiv \ket{11}$ with $E_{\rm I}=J\Delta$ (here the 1st $0$ or $1$ in the ket/bra denotes the state on the upper leg, the 2nd on the lower leg). Written in the rung eigenbasis, labeled by 4 letters ${\rm \{S,T,O,I\}}$, the leg nearest-neighbor Hamiltonian $h_{i,i+1}^{||}$ has two kinds of terms. One that represent hopping, and the other that cause ``scattering''. The hopping terms are
\begin{equation}
\label{eq:hopp}
\ket{\rm OS} \leftrightarrow \ket{\rm SO}, \, \ket{\rm IS} \leftrightarrow \ket{\rm SI},\, \ket{\rm OT} \leftrightarrow \ket{\rm TO},\, \ket{\rm IT} \leftrightarrow \ket{\rm TI}
\end{equation}
where the notation $\leftrightarrow$ denotes a mapping under $\frac{1}{2}h^{||}_{i,i+1}$, for instance, $h^{||}_{i,i+1} \ket{\rm OS}=2\ket{\rm SO}$, as well as $h^{||}_{i,i+1} \ket{\rm SO}=2\ket{\rm OS}$. Terms in Eq.(\ref{eq:hopp}) are called ``hopping'' because they exchange the two rung states involved. The scattering terms, on the other hand, change the rung states and are,
\begin{eqnarray}
\label{eq:scatt}
h^{||}_{i,i+1} (\ket{\rm TT}-\ket{\rm SS})=4(\ket{\rm OI}+\ket{\rm IO}),\\
h^{||}_{i,i+1} (\ket{\rm OI}+\ket{\rm IO})=4(\ket{\rm TT}-\ket{\rm SS}).\nonumber
\end{eqnarray}
All other terms not listed in Eqs.~(\ref{eq:hopp}) and~(\ref{eq:scatt}) map to zero. Explicitly, 
\begin{eqnarray}
\label{eq:zero}
&h^{||}_{i,i+1}& \ket{\rm OO}=0,\quad h^{||}_{i,i+1} \ket{\rm I\,I}=0,\quad h^{||}_{i,i+1} \ket{\rm ST}=0,\nonumber \\ 
&h^{||}_{i,i+1}& \ket{\rm TS}=0,\quad h^{||}_{i,i+1} (\ket{\rm OI}-\ket{\rm IO})=0,
\end{eqnarray}
as well as $h^{||}_{i,i+1} (\ket{\rm TT}+\ket{\rm SS})=0$. We note that all direct products of $\ket{\rm T}, \ket{\rm S},\ket{\rm O}$ and $\ket{\rm I}$ are eigenstates of the whole rung Hamiltonian $\sum_{i=1}^{L} h^{\perp}_{i}$, in particular, all two-rung states annihilated by the leg Hamiltonian, i.e.,~those given in Eq.~(\ref{eq:zero}), are eigenstates of $h^{\perp}_i+h^{\perp}_{i+1}$ (while for finite $J$ the $\ket{\rm TT}+\ket{\rm SS}$ is not). Therefore, if one has a product state in which all nearest-neighbor terms are annihilated by the leg Hamiltonian Eq.~(\ref{eq:zero}), or correspond to hopping Eq.~(\ref{eq:hopp}), $H$ will preserve the number of each of the 4 letters. Such states therefore form an invariant subspace that can be labeled by a number of each of the 4 letters that the states contain.

Let us list some of the simpler invariant subspaces. The simplest is a product state with an interchanging $\ket{\rm S}$ and $\ket{\rm T}$ states, i.e., either $\ket{\rm STST\ldots}$ or $\ket{\rm TSTS\ldots}$. These two states, called an ${\rm S-T}$ background, are eigenstates of the XX ladder, Eq.~(\ref{eq:XXladder}), with the eigenenergy $2(N_{\rm T}-N_{\rm S})J-\Delta J L$, where $N_{\rm T,S}$ is the number of the respective letters. From those two states one can immediately get a series of invariant subspaces by inserting into an ${\rm S-T}$ ``background'' a number of ${\rm O}$ or ${\rm I}$ states. For instance, $L$ states of the form $\ket{\rm STIST\ldots}$, where one ${\rm I}$ is at any of $L$ possible positions, form an invariant subspace. Using the hopping property of the leg Hamiltonian (\ref{eq:hopp}) we can see that when $H$ acts on such a state it will either move the ${\rm I}$ to one of the two neighboring sites, or leave it at its original place. A completely analogous thing happens with two (or more) inserted ${\rm I}$s in the ${\rm S-T}$ background. The dimension of such an invariant subspace is $L \choose N_{\rm I}$, where $N_{\rm I}$ is the number of ${\rm I}$s, while the total magnetization is $Z=-2 N_{\rm I}$. We can see that within one such subspace, characterized by a fixed number of ${\rm I}$s, the dynamics is of the same hopping kind as in a single XX chain. It is therefore immediately clear that the transport of magnetization is ballistic within such an invariant subspace. The speed of a propagation front can also be immediately read out and is $v=4$ in the units chosen. Using $\ket{\rm O}$ instead of $\ket{\rm I}$ in the above construction results in invariant ballistic subspaces with magnetization $Z=+2N_{\rm O}$, where $N_{\rm O}$ is the number of inserted ${\rm O}$s. The above construction gave us ballistic subspaces with any nonzero magnetization. There exists though also a ballistic subspace with zero total magnetization. To see this we note that if we insert a single two-rung state $(\ket{\rm OI}-\ket{\rm IO})$ into an ${\rm S-T}$ background, the $H$ acting on such a state will move ${\rm I}$ and ${\rm O}$ to the left and to the right, one site after each application, while it will preserve the number of each of the four rung letters. Such states, which are a product of an ${\rm S-T}$ background and $(\ket{\rm OI}_{j,k}-\ket{\rm IO}_{j,k})$ at the $j$-th and $k$-th rung therefore form an invariant subspace of dimension $L \choose 2$~\cite{foot1}, again with ballistic dynamics, but having total magnetization $0$. There are other, more complicated invariant subspaces that we shall not consider in the present work~\cite{foot2}. Note that the total dimension of invariant subspaces discussed in the present work grows at least as $\sim 2^L$ (their total relative size though goes to zero as $\sim 2^L/2^{2L}$ when $L \to \infty$).

Considering a plethora of invariant subspaces one might be tempted to think that the XX ladder is an integrable system. While the limits $J=0$ and the Hubbard limit are indeed integrable, for general $J$ and $\Delta$ the model is not integrable. It in fact displays a standard feature of chaotic quantum systems, namely  a random-matrix-like repulsion between nearest-neighbor eigenlevels~\cite{Haake}.
\begin{figure}[ht!]
\centerline{\includegraphics[width=0.35\textwidth]{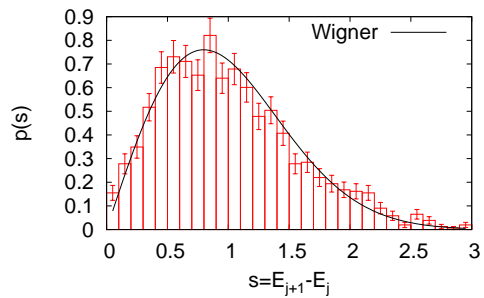}}
\caption{Level spacing statistics in the XX ladder (bars) agrees well with the Wigner's surmise (full curve), being a standard signature of quantum chaotic systems. Parameters are $J=1$, $\Delta=0$, $L=10$ (similar statistics is obtained also for $\Delta=1.5$) and the subspace with $Z=12$ and eigenvalues in the energy interval $E\in [1,9]$ are used.}
\label{fig:lsd}
\end{figure}
Calculating eigenenergies of the XX ladder and removing eigenenergies corresponding to the above-mentioned symmetries, the nearest-neighbor level spacing is well described (Fig.~\ref{fig:lsd}) by the Wigner's surmise, $p(s)=\frac{\pi}{2} s \exp{(-s^2 \pi/4)}$, that represents a good approximation to level spacing statistics in systems with orthogonal symmetry (the so-called Gaussian orthogonal ensemble)~\cite{Haake}. 

We have therefore seen that in the XX ladder, even-though it is in general chaotic, there is a ballistic subspace having any total magnetization. Ballistic transport, usually associated with local conserved quantities, has been so far observed only in integrable systems. Here, however, the invariant subspace is not associated with any local conserved quantity~\cite{foot21}. All our statements on the existence of invariant ballistic subspaces hold for any value of the rung anisotropy $\Delta$ as well as for any $J$ (in fact, one can also allow for an inhomogeneous rung as well as leg coupling strength). As a simple consequence, there exist ballistic subsectors also in the one-dimensional Hubbard model at any filling, including at half-filling for which diffusive transport has been observed~\cite{1dhubbard} (there is no contradiction because, as we shall see, the actual observed transport depends on details of the coupling with a bath). A heuristic picture of how ballistic transport comes about is quite simple: provided we have the right background state that does not interact with the ${\rm I}$s -- we can consider ${\rm I}$s as being elementary excitations -- the ${\rm I}$s move around freely. The same holds if only ${\rm O}$s are present. However, as soon as we leave the invariant subspace, by allowing both ${\rm I}$s and ${\rm O}$s, an ${\rm I}$ can collide with an ${\rm O}$, resulting in a ``scattering'' event that modifies the background, effectively introducing other types of excitation (a neighboring ${\rm T-T}$ or an ${\rm S-S}$). An interesting question is, what is the nature of transport in the subspace orthogonal to all ballistic ones? While we are not able to analytically answer this question we shall use large-scale numerical simulations to demonstrate that it is diffusive.

There are two ways to assess transport properties. (i) Coupling the system with reservoirs and studying a genuine nonequilibrium steady state, by, for instance, measuring the stationary current through the system. One possibility is to implement the Lindblad master equation with cold atoms. Essential ingredients have already been experimentally demonstrated~\cite{Diehl09}. (ii) Another possibility is without any external coupling to reservoirs, by simply preparing a nonequilibrium initial state, and then observing how the initial disturbance spreads. Such experiments have been performed~\cite{Zweirlein11,Schneider12} as well as appropriate numerical simulations~\cite{numdens}. For our ladder system one could prepare initial states from the ballistic subspace by quenching a strong dimerized rung coupling strength, so that the ground state would be $\ket{{\rm STS}\ldots}$, as well as strong local magnetic field introducing  ${\rm I}$ or ${\rm O}$ ``excitations''.

{\em Numerical demonstration.--} 
Here we focus on numerically demonstrating (we set $J=1$) the above findings via a master equation of the Lindblad form~\cite{Lindblad}, in which the action of reservoirs is described by a dissipator ${\cal L}^{\rm dis}$,
\begin{equation}
{{\rm d}}\rho/{{\rm d}t}=\ii [ \rho,H ]+ {\cal L}^{\rm dis}(\rho)={\cal L}(\rho).
\label{eq:Lin}
\end{equation}
The dissipator ${\cal L}^{\rm dis}$ is expressed in terms of Lindblad operators $L_k$, ${\cal L}^{\rm dis}(\rho)=\sum_k \left( [ L_k \rho,L_k^\dagger ]+[ L_k,\rho L_k^{\dagger} ] \right)$. After long time~\cite{foot3} the solution of the Lindblad master equation (\ref{eq:Lin}) converges to a nonequilibrium stationary state (NESS) $\rho_\infty=\lim_{t \to \infty} \rho(t)$. In all cases studied here the NESS is unique and therefore independent of the initial state. We are, in particular, interested in the NESS expectation value of the magnetization current $J$ in the upper leg, $J=\tr{(\rho_\infty j^{(\sigma)}_k)}$, $j^{(\sigma)}_k={\rm i}[\sigma_k^{\rm z},h^{||}_{k,k+1}]=2(\sigma_k^{\rm x}\sigma_{k+1}^{\rm y}-\sigma_k^{\rm y}\sigma_{k+1}^{\rm x})$. A similar expression can be defined for the lower leg. Crucial question then is how the current $J$ scales with $L$, provided we keep the driving constant. If one has $J \sim 1/L$, the transport is diffusive, if $J \sim L^0$, it is ballistic. We shall demonstrate that in the XX ladder, depending on the choice of Lindblad operators, one can have either ballistic or diffusive behavior.

\begin{figure}[ht!]
\centerline{\includegraphics[width=0.38\textwidth]{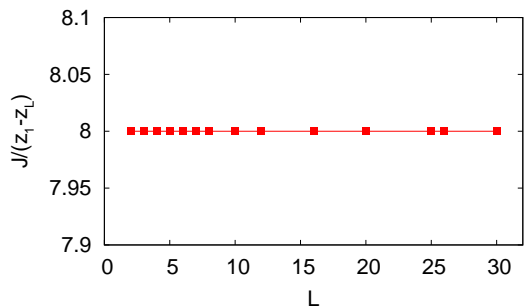}}
\caption{Ballistic transport in the zero and single ${\rm I}$-excitation subspace of the XX ladder, $\Delta=0$. Scaled current $J/(z_1-z_L)$ is independent of the system length $L$.}
\label{fig:conductance}
\end{figure}

{\em Ballistic case.--} Here we want to demonstrate ballistic transport in one of the invariant subspaces. We choose the subspace of dimension $2(L+1)$ spanned by two background states and $2L$ states containing one ${\rm I}$ in an ${\rm S-T}$ background. The idea is to use such Lindblad operators that inject ${\rm I}$ at the left end and remove it at the right end, while at the same time preserving the invariant subspace. One possible choice of Lindblad operators is $L_1=\ket{\rm ITS\ldots ST}\bra{\rm STS\ldots ST}$, $L_2=\ket{\rm IST\ldots TS}\bra{\rm TST\ldots TS}$, $L_3=\ket{\rm STS\ldots ST}\bra{\rm STS\ldots SI}$ and $L_4=\ket{\rm TST\ldots TS}\bra{\rm TS\ldots TI}$ (written for even $L$). $L_{1,2}$ inject one ${\rm I}$ while $L_{3,4}$ absorb one ${\rm I}$. They therefore preserve a union of zero and one ${\rm I}$-excitation subspaces. It should be clear from our explicit exposition in the first part of the Letter that within this subspace the dynamics should be ballistic. We have diagonalized the superoperator corresponding to the Lindblad equation (\ref{eq:Lin}) with the chosen four Lindblad operators. Finding NESS $\rho_\infty$ for $L$ up to $30$, we can see in Fig.~\ref{fig:conductance} that the current $J$ is indeed independent of the system size. The magnetization profile is, as expected for a ballistic system, flat in the bulk (magnetization is the same at 1st $L-1$ sites and different at the last; data not shown).
In fact, a very simple formula for the current holds, $J= \frac{v^2}{2}\, (z_1-z_L)$, where $v=4$ is the speed given by the bandwidth due to the hopping (\ref{eq:hopp}) and $z_k=\tr{(\rho_\infty \sigma_k^{\rm z})}$. Because of the symmetric driving currents and magnetization are the same in both legs.

\begin{figure}[ht!]
\centerline{\includegraphics[width=0.40\textwidth]{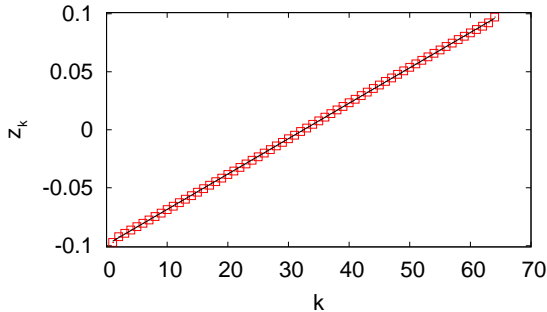}}
\caption{Magnetization profile along the ladder in the NESS for the symmetric driving (a) with 8 Lindblad operators. $L=64$, $\mu=0.1$, $\Delta=0$.}
\label{fig:profil}
\end{figure}

\begin{figure}[ht!]
\centerline{\includegraphics[width=0.45\textwidth]{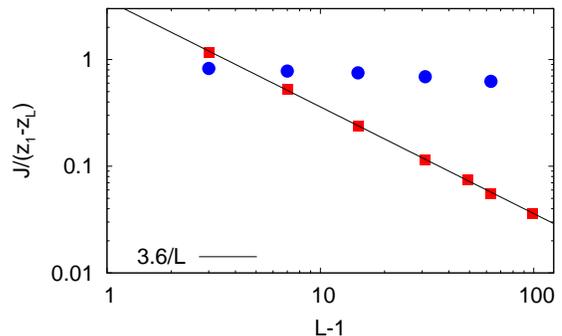}}
\caption{Scaling of the current with the system size $L$ for the driving with 8 Lindblad operators. Squares are for the symmetric driving with $\mu=0.2$ and $\Delta=0$, circles for the asymmetric one with $\Delta=1.5$ (in this case we plot $J/(\mu_L-\mu_1)=J/0.4$). The full line suggests diffusive $\sim 1/L$ scaling.}
\label{fig:jXX}
\end{figure}

{\em Local Lindblad operators.--} Here we will take eight Lindblad operators that will induce a nonequilibrium situation. They are $L_{1,2}=\sqrt{1\pm \mu_{1\sigma}}\sigma^\pm_1$ at the left end, $L_{3,4} =\sqrt{1\pm \mu_{L\sigma}}\sigma^\pm_L$ at the right end, and, similarly, $L_{5,6}=\sqrt{1\pm\mu_{1\tau}}\tau^\pm_1$, $L_{7,8} =\sqrt{1\pm\mu_{L\tau}}\tau^\pm_L$. Driving parameters $\mu$ try to induce a nonzero magnetization at the respective ladder site. $L_k$ do not preserve any of the invariant ballistic subspaces. We shall consider two cases: (a) symmetric driving around $z=0$, $-\mu_{1\sigma}=-\mu_{1\tau}=\mu_{L\sigma}=\mu_{L\tau}=\mu$, and (b) an asymmetric driving inducing a nonzero total magnetization, $\mu_{1\sigma}=0.5, \mu_{L\sigma}=0.9, \mu_{1\tau}=0.4, \mu_{L\tau}=0.8$. In the case (a) all contributions to the current from invariant ballistic subspaces mutually cancel and there is no net ballistic current (if there is a contribution from a subspace ${\cal S}$, there is also one from a spin-flipped counterpart $F({\cal S})$, in which all ${\rm O}$s are replaced by ${\rm I}$s, and which carries an opposite current). In the case (b) there is no such cancellation and there is a nonzero net ballistic contribution. For finite $L$, one will therefore have a combination of ballistic and diffusive contributions, with the ballistic one dominating long-time behavior. In (a) the NESS is, for small driving $\mu$, close to the identity matrix (for $\mu=0$ the NESS is $\rho_\infty \propto \mathbbm{1}$). For the asymmetric Lindblad operators, and if $\mu_{1\sigma}=\mu_{L\sigma}=\mu_{1\tau}=\mu_{L\tau}=\bar{\mu}$, the NESS is~\cite{foot25} $\rho_\infty \propto \prod_{j=1}^L (\mathbbm{1}+\bar{\mu}\sigma_j^{\rm z})(\mathbbm{1}+\bar{\mu}\tau_j^{\rm z})$. Therefore, comparing NESS states with the grandcanonical one, $\rho \propto \exp{(-\beta(H-\phi Z))}$, we can say that the NESS is for small driving in (a) close to the infinite temperature state, while in (b) it is close to an infinite temperature but finite $\beta \phi$ state. Different transport behaviors seen, e.g., in Fig.~\ref{fig:jXX} are therefore exhibited close to thermal states. There is in principle no difficulty in using the more complicated Lindblad operators leading to states at finite temperatures (see, e.g., 2nd Ref. under~\cite{Affleck} for an example) for which similar results are expected. For small driving NESS is therefore close to separable in the operator space, and one of the best methods for solving the Lindblad equation is a time-dependent density matrix renormalization group method in the operator space. Details of our implementation can be found in Ref.~\cite{tdmrg}. We managed to calculate NESS for systems of up to $L=100$ sites. In Fig.~\ref{fig:profil} we show a magnetization profile along one of the legs (it is the same in both) for the symmetric driving.  
We note that the linear profile is typical for diffusive systems. That the transport is indeed diffusive is confirmed by scaling of the current $J$ with the system size (Fig.~\ref{fig:jXX}, squares), showing a nice diffusive $J \approx \frac{3.6}{L} (z_1-z_L)$ relation. On the other hand, for the asymmetric driving the scaling of the current with the system size is markedly different -- the ballistic contribution is nonzero and dominant, circles in Fig.~\ref{fig:jXX}. An important question is, how stable are the phenomena presented; for instance, can one still have ballistic transport after an addition of small perturbation to $H$ given by Eq.~(\ref{eq:XXladder})? While a detailed answer goes beyond the present work, let us just mention that small interaction along legs of the form $J_z (\sigma_i^{\rm z} \sigma_{i+1}^{\rm z}+\tau_i^{\rm z} \tau_{i+1}^{\rm z})$ has very little influence on the current. Concretely, choosing $J_z=0.1$ and the asymmetric driving current changes by about $\sim 5\,\%$ and would be almost indistinguishable from the data for $J_z=0$ (circles in Fig.~\ref{fig:jXX}).  

{\em Conclusion.--} We have identified a number of ballistic invariant subspaces in a class of XX spin ladders. Such systems are in general quantum chaotic, with integrable limits, one of them being the one-dimensional Hubbard model, appearing at special parameter values. We therefore explicitly show that one can have ballistic transport in a chaotic system. Using extensive numerical simulations we in addition show that, outside of these invariant subspaces, the transport is diffusive. In a simple strongly correlated system one can therefore observe either ballistic or diffusive transport. Such behavior can be observed either in the time-evolution of inhomogeneous initial states, or in an open-system with external driving. Which of the two regimes is exhibited depends on the initial state or, in a master equation setting, on the symmetry of a particular driving. 
Support by the research Program P1-0044 is acknowledged.


\begin{thebibliography}{}
\bibitem{foot1} Dimensions of invariant subspaces that we list are without taking into account the parity $P_{\rm x}$ or the spin-flip in the $Z=0$ subsector.
\bibitem{foot2} For instance, all states that have one $\ket{{\rm I}}$ as well as one $\ket{\rm OI}_{j,k}-\ket{\rm IO}_{j,k}$ at rungs $j$ and $k$ in an ${\rm S-T}$ background again form an invariant subspace.
\bibitem{foot21} We have explicitly checked that there are no conserved quantities with its density supported on less than 4 consecutive rungs (apart from the energy and magnetization). In view of the level repulsion (Fig.~\ref{fig:lsd}) the existence of local higher-order conserved quantities seems unlikely.
\bibitem{foot3} The convergence time to $\rho_\infty$ is given by the inverse of the spectral gap of ${\cal L}$, Eq.~(\ref{eq:Lin}). The spectral gap scales with the system size as $\sim 1/L^2$ for the diffusive case (8 Lindblad operators), while it goes as $\sim 1/L^3$ in the ballistic case (4 Lindblad operators).
\end{thebibliography}
\end{document}